\documentclass{article}

\usepackage{PRIMEarxiv  }

\usepackage[utf8]{inputenc} 
\usepackage[T1]{fontenc}    
\usepackage{hyperref}       
\usepackage{url}            
\usepackage{booktabs}       
\usepackage{amsfonts}       
\usepackage{nicefrac}       
\usepackage{microtype}      
\usepackage{lipsum}
\usepackage{fancyhdr}       
\usepackage{graphicx}       
\usepackage{placeins}
\usepackage{xcolor}

\usepackage[
backend=biber,
style=phys,
sorting=none,
doi=false,isbn=false,
url=false,eprint=false,
date=year,
maxbibnames= 6
]{biblatex}
\usepackage{authblk}
\usepackage{subcaption}
\usepackage{amsmath}

\title{Digital reconstruction of squeezed light for quantum information processing}
\author[1]{Huy Q. Nguyen}
\author[1,2]{Ivan Derkach}
\author[1]{Adnan A.E. Hajomer}
\author[1,3]{Hou-Man Chin}
\author[2]{Akash nag Oruganti}
\author[1]{Ulrik L. Andersen}
\author[2]{Vladyslav Usenko}
\author[1,*]{Tobias Gehring}

\affil[1]{Center for Macroscopic Quantum States (bigQ), Department of Physics, Technical University of Denmark}
\affil[2]{Department of Optics, Faculty of Science, Palacky University, Olomouc, Czech Republic}
\affil[3]{Machine Learning in Photonic Systems, Department of Electrical and Photonic Engineering, Technical University of Denmark}
\affil[*]{tobias.gehring@fysik.dtu.dk}

\begin{document}
\maketitle
\begin{abstract}

Squeezed light plays a vital role in quantum information processing. By nature, it is highly sensitive, which presents significant practical challenges, particularly in remote detection, traditionally requiring complex systems such as active phase locking, clock synchronization, and polarization control. Here, we propose and demonstrate an asynchronous detection method for squeezed light that eliminates the need for these complex systems. By employing radio-frequency heterodyne detection with a locally generated local oscillator and applying a series of digital unitary transformations, we successfully reconstruct squeezed states of light. We validate the feasibility of our approach in two key applications: the distribution of squeezed light over a 10 km fiber channel, and secure quantum key distribution between two labs connected via deployed fiber based on continuous variables using squeezed vacuum states without active modulation. This demonstrates a practical digital reconstruction method for squeezed light, opening new avenues for practical distributed quantum sensing networks and high-performance and long-distance quantum communication using squeezed states and standard telecom technology.

\end{abstract}

\section{Introduction}

Quadrature-squeezed states are quantum states of light characterized by reduced noise in one quadrature compared to vacuum states or shot noise. Its non-classical nature has enabled a wide range of applications, from measurement-based quantum computing~\cite{menicucciUniversal2006,weedbrookGaussian2012,larsenDeterministic2021} to quantum metrology \cite{aasiEnhanced2013}, to quantum sensing~\cite{guoDistributed2020,nielsenDeterministic2023}. In particular, squeezed light has proven to be highly valuable in the realm of quantum communication and quantum key distribution (QKD). It has been used to enhance noise tolerance~\cite{madsenContinuous2012,garcia-patronContinuousVariable2009,usenkoSqueezedstate2011,derkachSqueezingenhanced2020}, facilitate composable and one-sided device-independent security~\cite{gehringImplementation2015}, and eliminate information leakage to adversaries~\cite{jacobsenComplete2018}. However, measuring squeezed light poses challenges due to stringent requirements on phase noise and mode matching, which has limited most prior quantum communication implementations to free-space or emulated loss channels using an optical attenuator~\cite{madsenContinuous2012,jacobsenComplete2018,gehringImplementation2015}.

Squeezed light is typically generated using nonlinear crystals via parametric down-conversion (PDC) and detected using coherent detection, a widely adopted technique in classical optical communication and quantum optics~\cite{andersen30Years2016,kikuchiFundamentals2016}. In this detection technique, the squeezed light is combined with a strong laser beam, referred to as the local oscillator (LO), on a balanced beam splitter, and the resulting signal is detected with a balanced photodetector (BD). The LO serves as a phase reference, a coherent amplifier, and a filter, selecting the signal mode~\cite{leonhardtMeasuring1997}. For efficient coherent detection, the LO mode must be precisely matched to the polarization, spatial, and temporal modes of the squeezed light while also maintaining a stable phase reference to accurately measure the squeezed quadrature.

To ensure optimal mode matching in coherent detection, the LO and squeezed light are often generated from the same laser source \cite{leonhardtMeasuring1997}. The relative phase between the two fields is then controlled with a phase shifter to select the squeezed quadrature. However, for practical applications, such as QKD, the generation and detection of squeezed light might occur at separate locations connected by an optical fiber or free-space channel. This creates additional challenges, as the LO beam at the detector must be phase-stabilized and mode-matched with the squeezed light. Conventionally, the relative phase between LO and squeezed light can be stabilized using a complex optical phase lock loop (OPLL) \cite{suleiman40KmFiber2022,hajomerSqueezed2024} or by multiplexing the LO with the squeezed light in the same communication channel \cite{chapmanTwomode2023}, which introduces a security vulnerability in QKD systems~\cite{huangQuantum2013,jouguetPreventing2013}. Moreover, random birefringence in the fiber channel causes the polarization of the squeezed light to drift, necessitating real-time polarization alignment with that of the LO to avoid any additional loss. This alignment is typically achieved using an analogue polarization controller. These effects caused by the fiber channel require significant efforts to stabilize phase and polarization in real-time to achieve stable, autonomous fiber-based quantum optics systems \cite{nakamura24hour2024}. Additionally, remote transmission of squeezed light also requires clock synchronization between the source and receiver, typically achieved by sharing the clock optically, which adds to system complexity~\cite{hajomerSqueezed2024}. These challenges motivate the development of a simpler method to reconstruct squeezed light post-measurement, allowing for digital compensation of channel-induced impairments.

In this work, we propose and experimentally demonstrate a method for reconstructing squeezed light post-measurement by applying a series of unitary transformations to digitally compensate for channel effects. Our approach relies on radio-frequency (RF) heterodyne detection, facilitated by a locally generated local oscillator (LLO), which enables simultaneous measurement of both field quadratures, making it particularly suitable for quantum communication applications \cite{lanceNoSwitching2005}. We validate the effectiveness of our method experimentally in two use cases. First, we implement a practical squeezed state distribution scheme over 10 km of single-mode fiber (SMF), which does not require phase locking and polarization alignment prior to measurement. Second, we implement a passive continuous-variable quantum key distribution system based on squeezed vacuum states, with squeezed light transmitted via deployed fiber to a remote location. This method significantly simplifies the distribution of squeezed light in fiber channels, paving the way for practical applications in quantum communication and distributed quantum sensing networks.

\section{RF heterodyne detection of squeezed light}
\label{sec:RF_het}
\subsection{Basic concept}
Squeezed light can be measured using coherent detection methods, such as homodyne detection, which measures a specific quadrature, or heterodyne detection, which captures both quadratures simultaneously, though with a 3 dB signal-to-noise ratio penalty. The latter, in particular, is widely used in quantum communication \cite{liContinuousvariable2014,jainPractical2022a,hajomerContinuousvariable2024}. Heterodyne detection can be implemented in two ways. One method, known as a phase-diversity receiver, splits the optical signal into two arms, each overlapping with an LO that is phase-shifted by 
$\frac{\pi}{2}$ in one arm. This allows two balanced detectors (BD) to measure both quadratures simultaneously. While this technique efficiently uses the detector's bandwidth, it is costly and prone to phase and amplitude imbalances due to device imperfections~\cite{pereiraImpact2021}.

An alternative solution is radio-frequency (RF) heterodyne detection, where both quadratures can be detected using a single BD by detuning the LO frequency relative to the signal. RF heterodyne detection is commonly adopted in optical communication and continuous-variable quantum key distribution (CV-QKD)~\cite{hajomerLongdistance2024,jainPractical2022a,chinDigital2022} due to its robustness against device imperfections like amplitude and phase imbalances, as well as its lower cost and simplified optical subsystem. However, to the best of our knowledge, it has not yet been employed for measuring squeezed light.

Quantum mechanically, the photocurrent from RF heterodyne detection can be expressed as~\cite{kikuchiFundamentals2016},
\begin{equation}
\begin{split}
    \hat{I} & \propto (\hat{a}_S^{\dagger} + \Delta \hat{a}_i) e^{-i \Delta \omega t} + (\hat{a}_S + \Delta \hat{a}_i^{\dagger}) e^{i \Delta \omega t} \\
    & \propto \hat{x}' \cos(\Delta \omega t) - \hat{p}'\sin(\Delta \omega t) ,
\end{split}
\label{eq:1}
\end{equation}
where $\hat{a}_S$ and $\Delta \hat{a}_i$ are the annihilation operators of the signal mode and the mode at the opposite frequency with respect to the LO (this is usually referred to as image band), $\Delta \omega$ is the frequency separation between the LO and the signal. Here, we can define two new effective quadrature operators:
\begin{equation}
    \begin{split}
        \hat{x}' &= \hat{a}_S^\dagger + \Delta \hat{a}_i + \hat{a}_S + \Delta \hat{a}_i^\dagger = \hat{x}_S + \Delta \hat{x}_i \\
        \hat{p}' &= i(\hat{a}_S^\dagger + \Delta \hat{a}_i - \hat{a}_S - \Delta \hat{a}_i^\dagger ) = \hat{p}_S - \Delta \hat{p}_i. \\
    \end{split}
\end{equation}
Unlike the normal quadrature operators, these new operators commute, i.e., $[\hat{x}',\hat{p}']=0$, allowing for their simultaneous measurement \cite{kikuchiFundamentals2016}. However, due to the image band's contribution ($\Delta\hat{x}_i, \Delta\hat{p}_i$), RF heterodyne detection, still incurs noise penalty~\cite{kikuchiFundamentals2016}. When detecting coherent states, this image band is usually considered to be a vacuum which entails a 3 dB noise penalty to the measurement. However, in the case of squeezed states, the image band is generally not a vacuum but an attenuated squeezed state. Therefore, careful attention must be given to the frequency separation between the LO and the squeezed signal, as will be discussed in the next section.

\subsection{Experimental implementation}

\begin{figure}[ht!]
    \centering
    \includegraphics[width=0.9\linewidth]{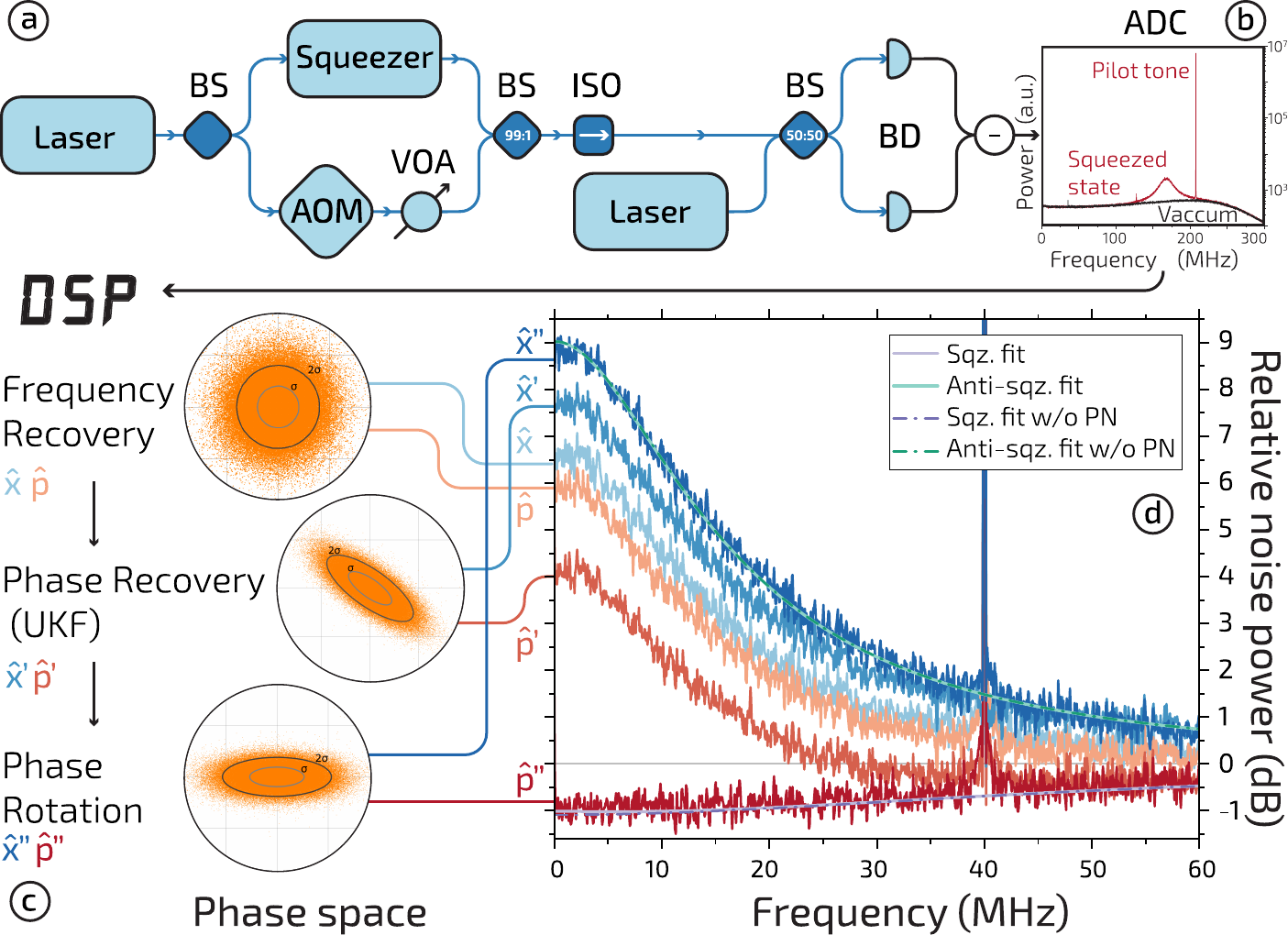}
    \caption{ \textbf{RF heterodyne detection of squeezed light}. (a) Schematic of the experimental setup, including key components: beam splitter (BS), acousto-optic modulator (AOM), variable optical attenuator (VOA), optical isolator (ISO), balanced detector (BD), analog-to-digital converter (ADC), and unscented Kalman filter (UKF). (b) Power spectrum of the ADC output signal. (c) Phase space representation of the squeezed state ensemble at different stages of digital signal processing (DSP). (d) Power spectral density normalized to vacuum noise after various DSP steps.}
    \label{fig:rf_het_scheme}
\end{figure}

The experimental setup implementing RF heterodyne detection for measuring squeezed light is shown in Fig.~\ref{fig:rf_het_scheme} (a). The squeezed light is generated via parametric down-conversion pumped with 775 nm light (for further details, see Ref.~\cite{arnbakCompact2019}). A strong coherent pilot tone at 40 MHz, essential for the carrier phase recovery, was generated by an acousto-optic modulator (AOM), and its power was adjusted by a variable optical attenuator (VOA). This pilot tone was then combined with the squeezed light on a 99:1 beam splitter. An optical isolator was placed in the path to prevent back reflections to the squeezed light source. 

At the receiver, RF heterodyne detection was performed by combining the squeezed light with the LO at a balanced beam splitter. The LO, generated from an independent and free-running laser, was frequency-detuned relative to the squeezed light. Finally, the electrical signal from the BD was captured and digitized using a 16-bit analog-to-digital converter (ADC) operating at a sampling rate of 1 GSample/s. 

To reconstruct the squeezed light post-measurement, we implemented the digital signal processing (DSP) pipeline illustrated in Fig.~\ref{fig:rf_het_scheme} (c). The process began with frequency recovery, where the frequency difference between the LO and the squeezed signal was estimated using the pilot tone. Specifically, this frequency difference was determined by performing a linear fit on the phase extracted from the pilot's analytic representation, which is obtained through the Hilbert transform. The next step was phase recovery. Instead of using a real-time OPLL~\cite{suleiman40KmFiber2022}, we employed an unscented Kalman filter (UKF) \cite{chinMachine2021} to estimate the fast phase drift between the LO and the squeezed light, enabling us to restore the phase reference for the squeezed light. Finally, a phase rotation step was performed to align the squeezed and anti-squeezed quadratures. This was achieved by exploiting the asymmetry of the squeezed states, allowing for the correct alignment of the quadratures through state rotation,
\begin{equation}
       \textbf{R}(-\Tilde{\Phi}) \left( \textbf{R}(\Phi) \gamma_r \textbf{R}^T(\Phi) \right)\textbf{R}^T(-\Tilde{\Phi}) = \\\begin{bmatrix}
       e^{-2r} + 2 \sinh(2r) \sin^2(\Tilde{\Phi} -\Phi) & -\sin(2 (\Tilde{\Phi} -\Phi)) \sinh(2r) \\
        -\sin(2 (\Tilde{\Phi} -\Phi)) \sinh(2r)  & e^{2r} - 2 \sinh(2r) \sin^2(\Tilde{\Phi} -\Phi)
    \end{bmatrix}\ ,
    \label{eq:squeezed_state_rotation}
\end{equation}
Here, $\gamma_r$ represents the covariance matrix of a pure single-mode squeezed state with the squeezing parameter $r$. The matrices $\textbf{R}(\Phi)$ and $\textbf{R}(\Tilde{\Phi})$ are rotation matrices, where $\Phi$ corresponds to the misalignment angle and $\Tilde{\Phi}$ represents the estimated rotation angle. As shown in Eq. \ref{eq:squeezed_state_rotation}, any misalignment (i.e. when $\Phi \neq \Tilde{\Phi}$ ) in the measured quadratures introduces correlations between the squeezed and anti-squeezed quadratures. Therefore, our method involved determining the rotation angle $\Tilde{\Phi}$ that minimizes this correlation, effectively correcting for the misalignment.

To illustrate the impact of each step in the DSP pipeline, Fig.~\ref{fig:rf_het_scheme} (c) and (d) show the phase space representation and the power spectrum of the recovered squeezed state at various stages. Each state in phase space was estimated across the bandwidth of 25 MHz and interpolated from 40 ADC samples. The initial power spectrum of the ADC output, depicted in Fig.~\ref{fig:rf_het_scheme} (b), reveals the two-sided spectrum of a squeezed state degraded by phase noise, alongside a pilot tone located near 200 MHz, indicating the frequency separation between the LO and the laser used to generate the squeezed state.  After frequency recovery, achieved by estimating the pilot tone's frequency, the measurement outcomes were demodulated into the quadrature components $X$ and $P$. The resulting power spectrum is shown as the faded red and blue traces in Fig.~\ref{fig:rf_het_scheme} (d). At this stage, the phase-space distribution of the squeezed state ensemble (shown at the top in Fig.~\ref{fig:rf_het_scheme}(c)) appears smeared due to phase noise. The next step, phase recovery, involves estimating and compensating for the fast phase drift. This phase correction increases the noise in the $X$ quadrature (light blue trace in figure~\ref{fig:rf_het_scheme} (d)) and reduces the noise in the $P$ quadrature (light red trace in figure~\ref{fig:rf_het_scheme} (d)). In phase space, the state now adopts the typical elliptic shape of a squeezed state, as shown in the middle of Fig.~\ref{fig:rf_het_scheme}(c). Finally, a phase rotation is applied to minimize the correlation between the $X$ and $P$ quadratures, successfully aligning the squeezed and anti-squeezed quadratures. The results of this final step are represented by the dark red and blue traces in Fig.~\ref{fig:rf_het_scheme} (d) and the corresponding phase-space representation in the bottom of Fig.~\ref{fig:rf_het_scheme} (c).

To confirm that the recovered states are indeed squeezed states, we fitted the theoretical model of squeezed states as described in \cite{arnbakCompact2019, collettSqueezing1984, suleiman40KmFiber2022} to the obtained noise power spectra of the squeezed and anti-squeezed quadratures, normalized to the vacuum noise. The solid lines and dot-dash lines in Fig.~\ref{fig:rf_het_scheme} (d) represent the fitted theoretical model for two recovered quadratures in the presence and absence of phase noise, respectively. This fit closely matches the relative power of the recovered quadratures, indicating not only the successful recovery of the squeezed states but also demonstrating the effectiveness of the digital phase compensation approach.

\begin{figure}[ht!]
    \centering
\includegraphics[width=\linewidth]{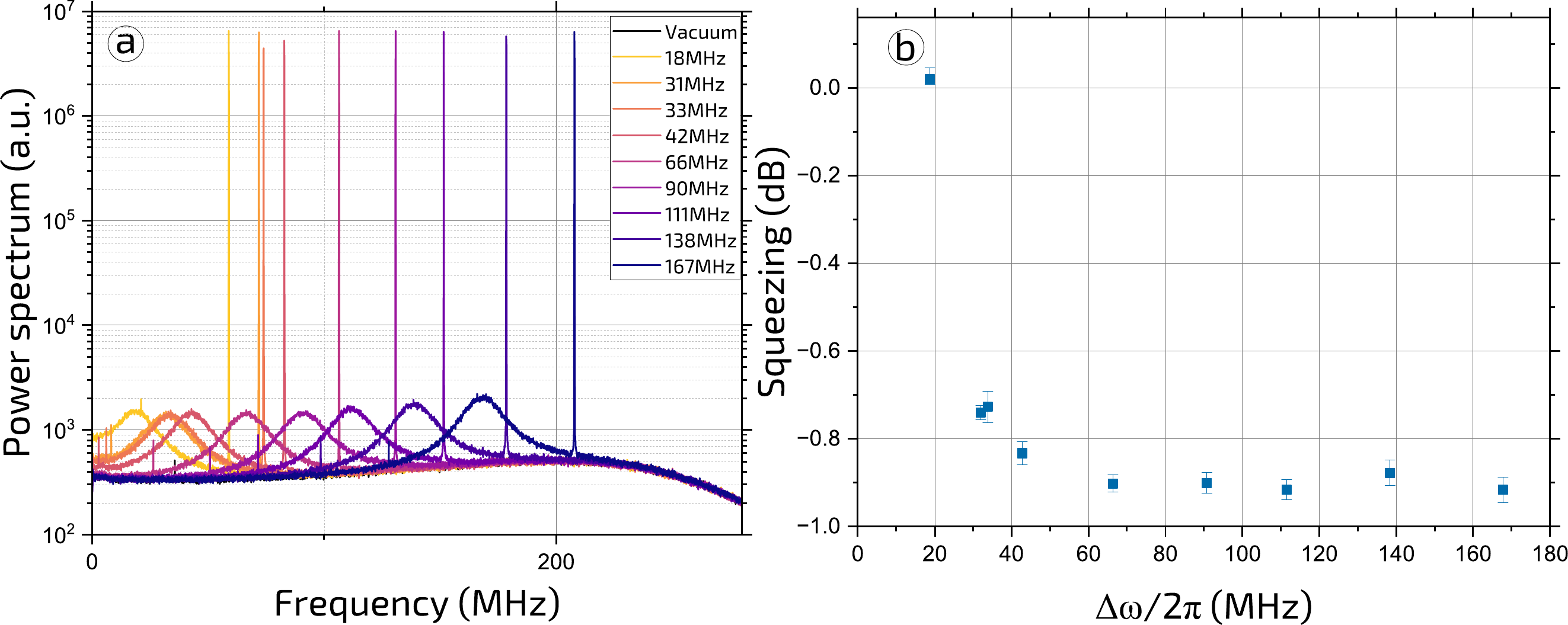}
\caption{ \textbf{Reconstruction of squeezed states with varying frequency separations}. (a) Power spectrum of ADC output at different LO frequencies. (b) Variance of the reconstructed squeezing quadratures across varying frequency separations.}
\label{fig:beat_shift_results}
\end{figure}
\FloatBarrier

As discussed previously, in heterodyne detection of squeezed states, unlike classical states, the image band might be non-vacuum. Understanding the penalty introduced by this non-vacuum image band is crucial. To explore this effect, we detuned the LO frequency so that the frequency difference ($\Delta\omega/2\pi$ in Eq.\ref{eq:1}) between the LO and the squeezed signal ranged from 18 MHz to 167 MHz, as shown in the spectrum in Fig.~\ref{fig:beat_shift_results} (a). Figure~\ref{fig:beat_shift_results} (b) shows that after applying our DSP chain, smaller frequency separations result in a greater penalty on the squeezing, as the power of the image band becomes significant, particularly at $\Delta\omega/2\pi = 18$ MHz no squeezing can be recovered. However, as the frequency separation increases, the penalty plateaus around 100 MHz. The degradation in squeezing arises from uncorrelated noise present in the image band. When the frequency separation is too narrow, significant power remains in the image band, this uncorrelated noise mixes with the squeezed signal during heterodyne detection and reduces the observed squeezing levels. However, the power of the image band attenuates rapidly at higher frequencies, so with sufficient frequency separation, the impact caused by the non-vacuum image band becomes negligible. These findings highlight the importance of optimizing the frequency separation relative to the squeezing bandwidth when performing RF heterodyne detection.

\section{Applications}
In this section, we present two practical applications of our proposed detection method and introduce additional DSP building blocks, aimed at simplifying the optical subsystem and enhancing practicality.  
\subsection{Practical squeezing distribution}
\label{sec:practical_sqz_dist}
Distributing squeezed light over optical fiber can enable applications like distributed quantum sensing networks~\cite{guoDistributed2020}. However, practical challenges -- primarily phase noise and polarization fluctuations -- must be addressed for effective squeezed state transmission over fiber channels. The phase noise issue can be mitigated using the proposed method described above. To tackle polarization fluctuations caused by random birefringence in the fiber channel, we introduce a digital polarization-diversity receiver designed for squeezed light reconstruction. Additionally, our system employs digital clock synchronization, eliminating the need for a common reference clock between remote stations and further enhancing system practicality.

\begin{figure}[b]
    \centering
    \includegraphics[width=\linewidth]{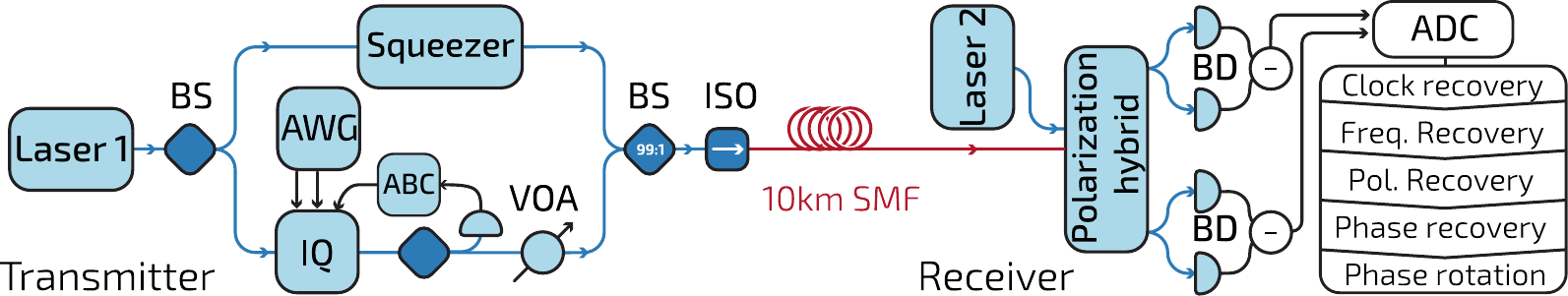}
    \caption{\textbf{Squeezed light distribution system}. IQ: in-phase and quadrature modulator; AWG: arbitrary waveform generator; ABC: automatic bias controller, BS: beam splitter, VOA: variable optical attenuator, ISO: optical isolator, BD: balanced detector, ADC: analog-to-digital converter.}
    \label{fig:sqz_states_transmit}
\end{figure}

Figure~\ref{fig:sqz_states_transmit} shows the schematic of the setup used for squeezed light distribution. This setup employed the same transmitter described in Section~\ref{sec:RF_het}, but with a key difference: instead of an AOM, we used an IQ modulator along with an arbitrary waveform generator (AWG) and an automatic bias controller (ABC) to generate two pilot tones that accompany the squeezed signal through the 10 km optical fiber channel. These pilot tones are used to facilitate clock and phase recovery. 

 At the receiver station, the optical subsystem included an independent, free-running laser to generate the LO, with sufficient frequency separation for RF heterodyne detection. A polarization-diversity optical hybrid was used to decompose the incoming light's polarization state into two orthogonal polarizations, which are then detected by two BDs and digitized using a two-channels ADC with a sampling rate of 1 GSample/s.
 
 The DSP module for reconstructing the squeezed light signal begins with a digital clock recovery block that utilizes the two pilot tones~\cite{chinDigital2022}. Given the independent clock systems at the transmitter and receiver, minor discrepancies arise in their respective frequency definitions (e.g., $f_{Pilot}^{Tx} \approx f_{Pilot}^{Rx}$). This discrepancy introduces errors in estimating the frequency offset between the squeezed state and the LO, potentially affecting system performance. However, by analyzing the frequency difference between the two pilot tones at the receiver, we can calculate a modifier, 
 $ \Delta f = \Delta f^{Rx}/\Delta f^{Tx}$,
to account for clock discrepancies and correct for drift \cite{chinDigital2022}. Here, $\Delta f^{Rx}$ and $\Delta f^{Tx}$ represents the frequency difference between the pilot tones at the receiver and transmitter respectively. Using this frequency modifier, along with the estimated pilot frequencies, we can accurately compensate for the frequency offset between the squeezed light and the LO. 

The next step in the DSP chain is polarization recovery, which addresses the random rotation of the squeezed light's state of polarization (SOP) due to random birefringence in the fiber channel. This polarization rotation can be represented as~\cite{savoryDigital2010},
\begin{equation}
    \begin{bmatrix}
        \textbf{E}_{rx}^{X} \\
        \textbf{E}_{rx}^{Y}
    \end{bmatrix} = 
    \begin{bmatrix}
        \cos(\theta) &  \sin(\theta) e^{-i \varphi} \\
        -\sin(\theta)  e^{-i \varphi} & \cos(\theta) 
    \end{bmatrix}
    \begin{bmatrix}
        \textbf{E}_{tx}^{X} \\
        \textbf{E}_{tx}^{Y}, \\
    \end{bmatrix}
    \label{eq:channel_model}
\end{equation}
where $\theta$ is the polarization rotation angle and $\varphi$ represents the additional phase shift introduced by the fiber channel, which will be compensated for in the subsequent DSP stage. To correct for this SOP rotation, we construct an inverse polarization rotation matrix by estimating the rotation angle that maximizes the power in the X polarization component while minimizing it in the Y polarization component, effectively shifting power between polarization components. Finally, we apply phase recovery and phase rotation, as detailed in the previous section, to reconstruct the squeezed state accurately.

\begin{figure}[ht!]
    \centering
  \includegraphics[width=\linewidth]{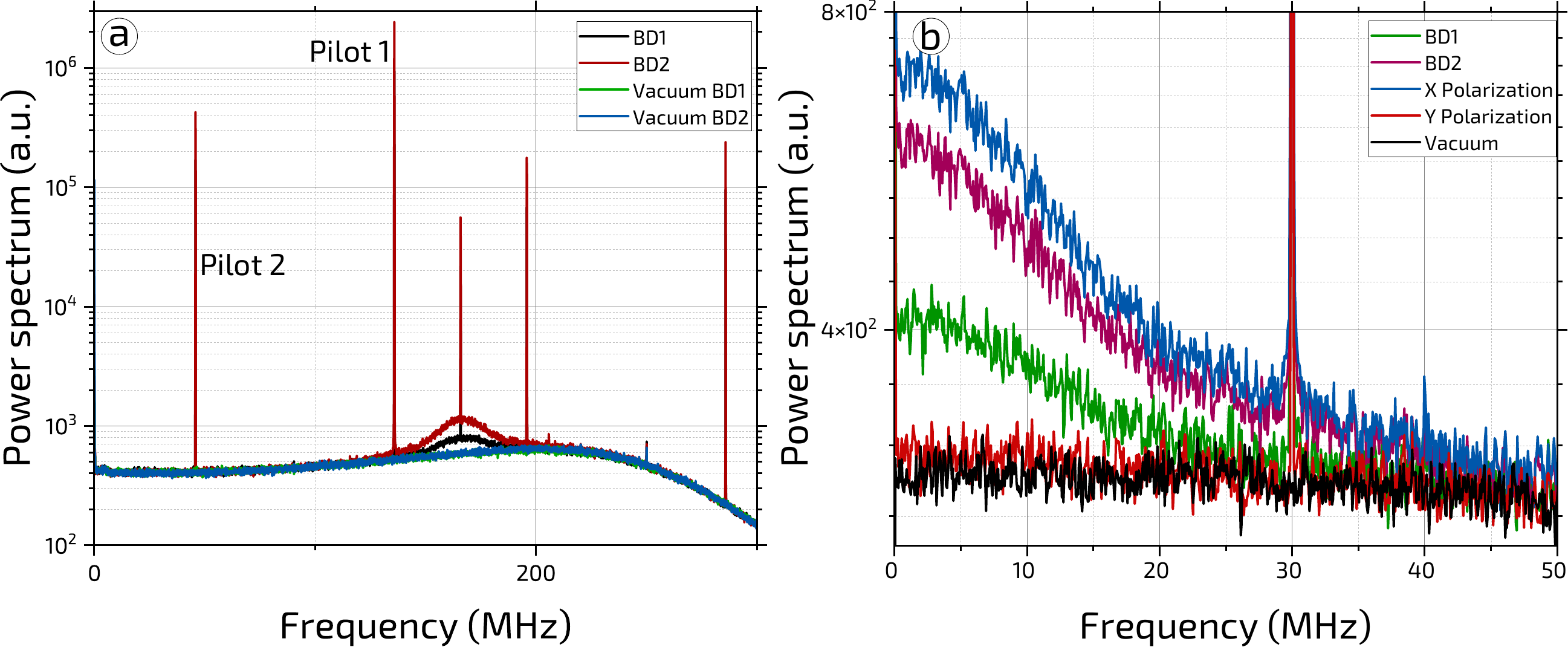}
\caption{ \textbf{Impact of digital polarization compensation}. (a) Power spectrum of detected optical signals (b) Power spectrum of signals after frequency and polarization recovery.}
\label{fig:PolCompPSD}
\end{figure}
Figure~\ref{fig:PolCompPSD} (a) and~\ref{fig:PolCompPSD} (b) show the spectra at the ADC outputs and after applying polarization recovery, respectively. The two pilot tones are modulated at 30 MHz and 120 MHz. Due to imperfect sideband suppression, two additional tones can be seen on the opposite side of the carrier. The Y polarization (red trace in Fig.~\ref{fig:PolCompPSD} (b)), closely matching the vacuum (black trace), indicates that most of the signal power has been successfully redirected into the X polarization, as compared to the spectrum traces after BD1 and BD2. This result demonstrates the effectiveness of our polarization compensation. With the polarization correctly aligned, we proceed with phase recovery and phase rotation, as detailed in section \ref{sec:RF_het}, to recover the squeezed and anti-squeezed quadratures of the state.

\begin{figure}[ht!]
\centering
    \centering
    \includegraphics[width=0.6\linewidth]{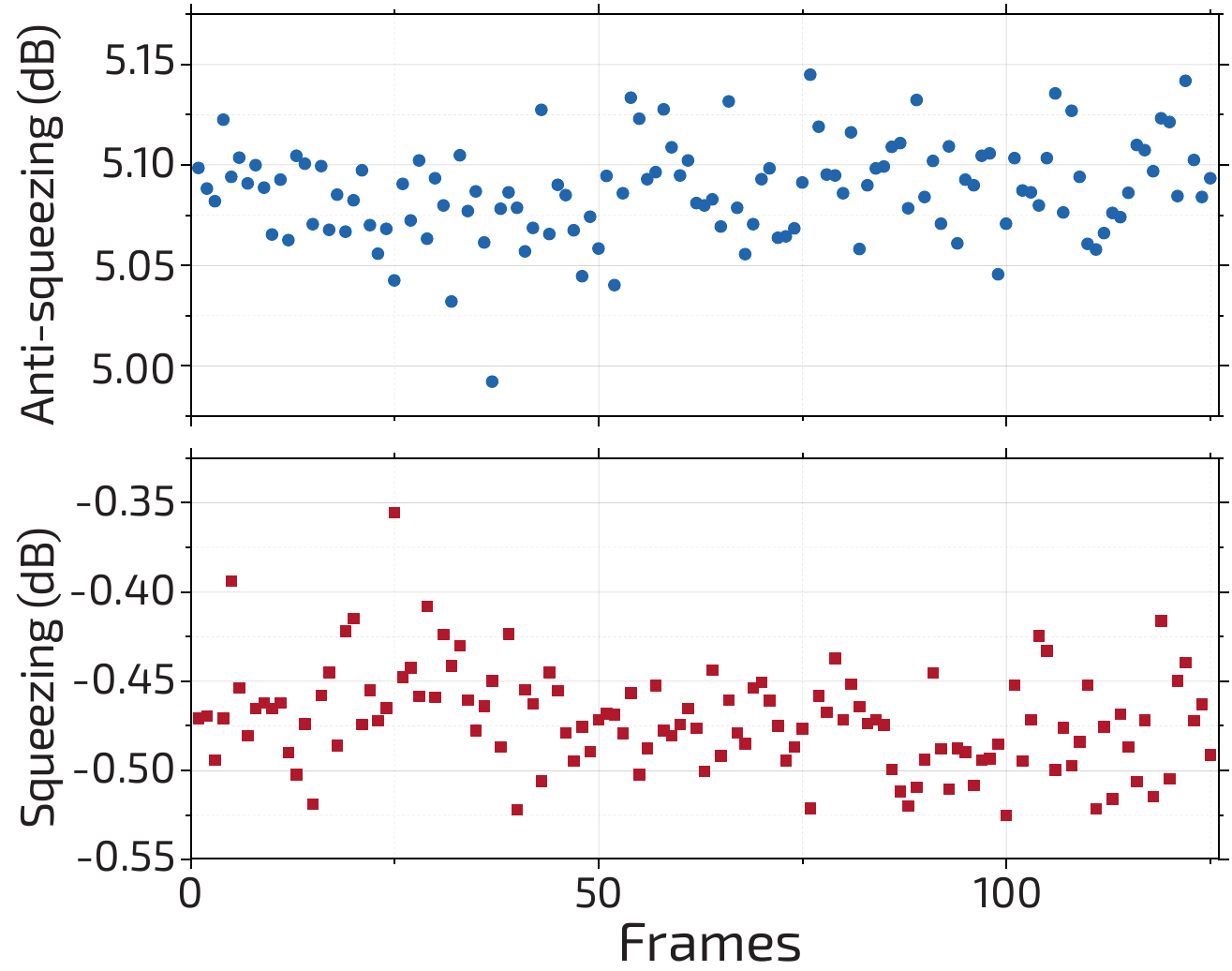}
    \caption{\textbf{Experimental results of squeezed distribution over  10 km fiber channel.} Squeezing and anti-squeezing levels across 125 frames, each frames consists of $2 \times 10^{5}$ states.}
    \label{fig:sqz-anti-sqz-results}  
\end{figure}

Figure \ref{fig:sqz-anti-sqz-results} presents the measured squeezing and anti-squeezing quadratures across a 25 MHz bandwidth. Each squeezed state was interpolated from ADC samples, with 40 samples per state. Analyzing the squeezing and anti-squeezing levels over 125 frames, each containing $2\times10^{5}$ states, we observed approximately -0.47 dB of squeezing and 5.07 dB of anti-squeezing. These results demonstrate the feasibility of practical squeezed light distribution using our proposed detection method.  

\FloatBarrier
\subsection{Passive CV-QKD with squeezed vacuum states over deployed fiber}
\label{sec:passive-CV-QK}
\begin{figure}[b]
    \centering
    \includegraphics[width=0.9\linewidth]{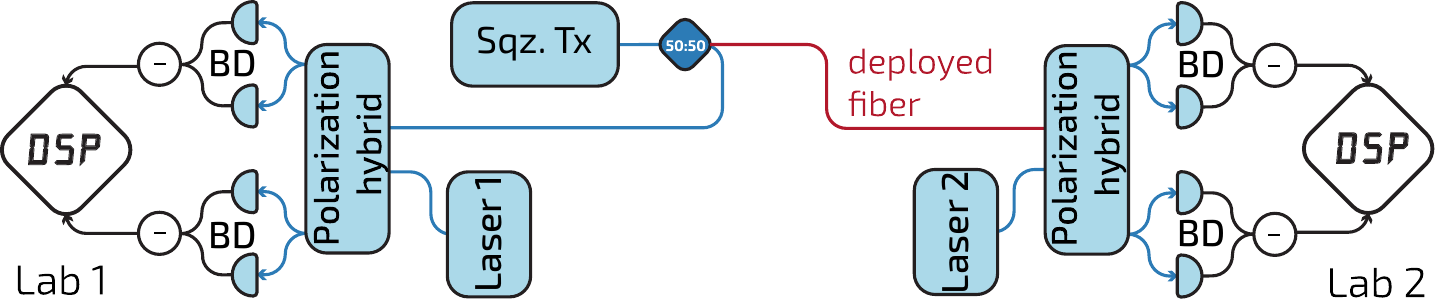}
    \caption{\textbf{Passive CV-QKD system}. The two labs are connected using a deployed fiber link. To establish the correlation between the two parties, the same DSP chain outlined in section \ref{sec:practical_sqz_dist} was applied to digitally reconstruct the squeezed state at each station.}
    \label{fig:experimental_method}
\end{figure}
Our method also has potential applications in quantum communication, particularly for distributing secure keys between two parties using CV-QKD protocols. Splitting a quadrature-squeezed mode leads to residual correlations between the output modes \cite{eberleGaussian2013}, and resulting impure squeezed states can be employed as foundation for a QKD protocol. In the following section we show how splitting of a single squeezed-light beam between two remote parties can lead to a variety of different secure keys, all based on residual
correlations between the measurement outputs. 

To show this, we implemented a passive CV-QKD system based on a single-mode squeezed vacuum state between two remote laboratories connected via SMF, as shown in Fig.~\ref{fig:experimental_method}. The squeezed-state transmitter is the same as described in the previous section. A balanced beam splitter splits the squeezed light between two stations: the local station (\textit{Lab 1}) and the remote station (\textit{Lab 2}), connected via the campus fiber network. Both stations use independent, free-running lasers as LOs and are equipped with polarization hybrid receivers. This setup requires neither prior polarization alignment, nor synchronization of the digital clocks to a common reference. Each station digitally reconstructs the squeezed state using our proposed DSP chain, detailed in the previous section.  Our system has a bandwidth of 25 MHz, corresponding to a symbol rate of 25 Mbaud.

The covariance matrix constructed based on parameter estimation (with detector electronic noise removed) with $2.4 \times 10^6$ quantum symbols reads (all parameters normalized to shot noise unit), 
\begin{equation} \label{eq:cov-matrix}
    \gamma =\begin{bmatrix}
    0.9149  \pm 0.0041 &  1.4365 \times 10^{-5} & 0.0836 \pm 0.0022 & 0.0028 \pm 0.0019 \\ 
   1.4365\times 10^{-5} &  2.6854 \pm 0.0144 &  0.0027 \pm 0.0019 & 1.5886 \pm 0.0095 \\
0.0836 \pm 0.0022      & 0.0027 \pm 0.0019  & 0.9346 \pm 0.0045 & 1.4379\times 10^{-5} \\
0.0028 \pm 0.0019	  &  1.5886 \pm 0.0095 &  1.4379\times 10^{-5} & 2.5297 \pm 0.0137
 \\
\end{bmatrix}.
\end{equation}
The channel loss, estimated by comparing the pilot power before and after transmission through the channel, was approximately -0.47 dB. The physical length of the fiber link is approximately 50 m. The extra loss is caused by several connectors patching different parts of the existing fiber network. By measuring the variance of electronic noise and characterizing detector efficiencies, we directly determined channel loss from the variances in the covariance matrix (\ref{eq:cov-matrix}). The loss value obtained using the pilot tone aligns with the loss estimated from the anti-squeezed quadrature variance, falling within the confidence interval of the latter. Based on the estimated channel losses, detector efficiencies, and electronic noises, we construct a purification model of the setup \cite{garcia-patronUnconditional2006a}. To eliminate numerical discrepancies between experimentally measured variances and theoretically expected quadrature variances, we assume thermal excess noise is present in the deployed fiber channel. Experimentally observed covariances between quadratures match theoretically expected ones with less than 1\% numerical discrepancy. The purification model of the setup attributes all anti-squeezing to trusted parties, i.e., assumes impurity of the squeezed state is due to the characterized and monitored signal source. Such an assumption was shown to partially reduce the impact of the noise in the squeezed-state CV-QKD protocol \cite{usenkoTrusted2016, orugantiContinuousvariable2024}. \par

The need for purification model stems from generation of impure squeezed states in practice. The signal state thus needs to be characterized in order for it to be used securely in a communication protocol. The purification model allows us to assess the quantum bosonic channel capacity $\chi(E)=S(E)-S(E|B)$, where $S(E)$ is the von Neumann entropy of the state of Eve $E$, and $S(E|B)$ is the conditional entropy, conditioned on the measurement of the reference side $B$, that upper limits the knowledge on the measurement results of the trusted reference side of the protocol. The Holevo bound $\chi(E)$ together with the readily available mutual information between the receivers $I(\text{Lab 1}:\text{Lab 2})$, that describes classical bosonic channel capacity between them, lower bounds the asymptotic key rate secure against Gaussian collective attacks \cite{navascuesOptimality2006, garcia-patronUnconditional2006a} as: $K\leq\text{max}\left[0,\beta\times I(\text{Lab 1}:\text{Lab 2}) - \chi(E)\right]$, where $\beta\in (0,1)$ is the error-correction efficiency, which defines how close the trusted parties can get to the classical channel capacity given by the mutual information. \par

The correlations in the squeezed $X$ and the anti-squeezed $P$ quadrature measurements will differ in strengths, and thus will lead to different levels of mutual information $I(\text{Lab 1}:\text{Lab 2})^X<I(\text{Lab 1}:\text{Lab 2})^P$. However, the eavesdropper, generally, will have more information on the anti-squeezed quadrature as well. Hence, either (or both) of the quadratures are viable for the key generation, and the final performance will depend on channel and source parameters. Therefore, there are two distinct approaches for Alice (Lab 1) and Bob (Lab 2) to obtain the secret key rate: 
\begin{enumerate}
    \item Using only measurements of a \textit{single quadrature}. The measurement results of complementary quadrature are thus used solely for monitoring and improvement of parameter estimation.
    \item Using \textit{both quadratures} for key generation (similarly to the coherent-state protocol \cite{weedbrookQuantum2004a}) with total mutual information now given as $I(\text{Lab 1}:\text{Lab 2})^X+I(\text{Lab 1}:\text{Lab 2})^P$. 
\end{enumerate}
In the former approach, the complementary quadrature does not contribute to the final key and thus can be announced publicly. The announcement can be used for a more accurate parameter estimation, and since conditioning does not increase the von Neumann entropy of the state available to Eve \cite{weedbrookGaussian2012}, it allows to raise the overall performance of the protocol (at least in the asymptotic regime). The mutual information between respective quadrature measurements will remain the same regardless of the chosen approach. However, the Holevo bound $\chi(E)$ will depend on whether one of the quadrature measurements is fully disclosed or not. For example if the $P$-quadrature measurement of Bob is completely disclosed the upper bound on Eve's knowledge of Bob's $X$-quadrature becomes $\chi(E|B^P)=S(E|B^P)-S(E|B^PB^X)$, which can be lower than the one conventionally used in the squeezed-state QKD protocols: $\chi(E|B^P)\leq\chi(E)$ \cite{cerfQuantum2001}. For further details on all relevant theoretical models and secure key analysis see Ref.\cite{orugantiContinuousvariable2024}. 

Either Alice or Bob can serve as a reference side for the error correction. However, generally, the party with lower loss is chosen to be the reference similarly as in Measurement Device-Independent protocols \cite{liContinuousvariable2014, pirandolaHighrate2015, hajomerHighrate2023}. In our case the receiver in \textit{Lab 1} was chosen to be the reference side, as shown in Fig.~\ref{fig:experimental_method}.

\begin{table}[]
\centering
\begin{tabular}{|l|l|l|l|}
\hline
\begin{tabular}[c]{@{}l@{}}Secure key rate\\ (in bits/symbol):\end{tabular}     & in X-quadrature ($K^X$) & in P-quadrature ($K^P$) & using both quadratures ($K^{XP}$) \\ \hline
\begin{tabular}[c]{@{}l@{}}Perfect reconciliation \\ $\beta=100$\% \end{tabular} & $2.49\times 10^{-4}$    & $1.34\times 10^{-3}$    & $9.36\times  10^{-4}$             \\ \hline
\begin{tabular}[c]{@{}l@{}}Realistic reconciliation\\ $\beta=95$\% \end{tabular} & $8.669\times 10^{-6}$   & 0                       & 0                                 \\ \hline
\end{tabular}
    \caption{Comparison of secure key rates obtained from different quadratures measurements.}
    \label{tab:keys}
\end{table}

In the asymptotic regime the keys secure against the collective attacks are summarized in Tab. \ref{tab:keys}. The key from the squeezed quadrature $K^X=\beta I(\text{Lab 1}:\text{Lab 2})^X-\chi(E|B^P)$ can be obtained with a rate of $K^X=2.49\times 10^{-4}$ bits/symbol. The mutual information between trusted stations is low due to small signal-to-noise ratio and weak correlations. However, the Holevo bound is similarly very low in the sub-shot-noise regime \cite{jacobsenComplete2018}, enabling a positive key rate. Under realistic reconciliation efficiency of $\beta=95\%$ ~\cite{maniMultiedgetype2021c} the rate reduces to $8.669\times 10^{-6}$ bits/symbol, yet remains positive. \par
For the anti-squeezed quadrature (with trusted additional anti-squeezing noise) both the mutual information $I(\text{Lab 1}:\text{Lab 2})^P$ and the Holevo bound $\chi(E|B^X)$ are larger, and the overall feasibility of positive key rate $K^P=\beta I(\text{Lab 1}:\text{Lab 2})^P-\chi(E|B^X)$ is more sensitive to the reconciliation efficiency, channel loss and noise \cite{orugantiContinuousvariable2024}. However, with the current parameters (assuming perfect efficiency $\beta=100\%$), we can obtain $K^P=1.34\times 10^{-3}$ bits/symbols of the secure key rate. Note that if additional anti-squeezing noise is attributed to the influence of the eavesdropper, the key generation in $P$-quadrature would not be feasible. 
Using all measurement results to generate the key as $K=\beta (I(\text{Lab 1}:\text{Lab 2})^X+I(\text{Lab 1}:\text{Lab 2})^P)-\chi(E)$ does not generally improve the overall rate \cite{garcia-patronQuantum2007, orugantiContinuousvariable2024}, as the conditional von Neumann entropy is now larger, and leads to an asymptotic rate of $9.36\times  10^{-4}$ bits/symbol. However, using both quadratures can prove to be beneficial in the finite-size regime with lower data block lengths and/or high anti-squeezing noise ~\cite{orugantiContinuousvariable2024}. Current methods for estimating security under finite-size effects apply only to CV QKD protocols with signal modulation \cite{ruppertLongdistance2014}, and a new theoretical approach would be required for a rigorous treatment of passive protocols in this regime. 

\section{Discussion}
Squeezed light is a critical resource for quantum information processing, yet its full potential in practical applications has been hindered by the complexity of detection methods ~\cite{suleiman40KmFiber2022, hajomerSqueezed2024, chapmanTwomode2023, nakamura24hour2024}. In this study, we proposed and demonstrated a practical approach for detecting squeezed light that simplifies measurements by eliminating the complex active locking systems required in traditional methods, thus enabling an asynchronous detection technique. 

Our approach introduces a digital coherent receiver for squeezed light detection, built on an RF heterodyne setup facilitated by an LLO and a carefully designed DSP chain to compensate for channel and measurement impairments. We demonstrated the feasibility of this digital receiver in two applications: first, squeezed light transmission over a fiber channel, supporting potential uses such as distributed quantum sensing networks; and second, passive CV-QKD over a deployed fiber channel between two laboratories. Compared to previous methods designed to address side-channel vulnerabilities at the transmitter using passive CV-QKD with thermal \cite{qiPassive2018}, coherent \cite{wuPassive2021, liPassive2022}, and squeezed states \cite{eberleGaussian2013, gehringImplementation2015}, our work represents the demonstration of passive CV-QKD over a deployed fiber channel. This proof-of-concept experiment underscores the technology's readiness for real-world deployment.

We believe our approach holds promise not only in point-to-point and multi-user network quantum communication but also in quantum sensing and computing, offering a foundation to harness the unique properties of squeezed light in practical settings and opening new possibilities for groundbreaking applications.

\section*{Acknowledgments}
The authors acknowledge support from the Danmarks Frie Forskningsfond (grant agreement no. 0171-00055B), from the Carlsberg Foundation (project CF21-0466), from the QuantERA II Programme (project CVSTAR) funded by EU Horizon 2020 research and innovation program (grant agreement no. 101017733), from Innovation Fund Denmark (CyberQ, grant agreement no. 3200-00035B), and the DNRF Center for Macroscopic Quantum States (bigQ, DNRF142). This project has received funding from the European Union’s Horizon Europe research and innovation programme under the project ``Quantum Security Networks Partnership'' (QSNP, grant agreement No 101114043). I. D. acknowledges the project 22-28254O of the Czech Science Foundation. V. U. acknowledges the project 21-44815L of the Czech Science Foundation and the project CZ.02.01.01/00/22$\_$008/0004649 (QUEENTEC) of the Czech MEYS. The data was processed using the resources from DTU Computing Center \cite{HPC_cite}.

\printbibliography

\end{document}